\documentclass[11pt]{article}

\usepackage{amsmath}
\usepackage{amssymb}
\def\beq{\begin{equation}}
\def\eeq{\end{equation}}

\newcommand{\ed}{\end{document}}

\begin{document}
\title{Observable Quantities in Weyl Gravity}

\author{M.R. Tanhayi$^1$\thanks{e-mail:
m$_{-}$tanhayi@iauctb.ac.ir} , M. Fathi$^1$\thanks{e-mail:
mohsen.fathi@gmail.com} , M.V. Takook$^2$\thanks{e-mail:
takook@razi.ac.ir}}

\maketitle   \centerline{\it $^{1}$Department of Physics, Islamic
Azad University, Central Tehran Branch, Tehran, Iran}
\centerline{\it$^2$Department of Physics, Razi university,
Kermanshah, Iran}
\begin{abstract}

In this paper, the cosmological ``constant" and the Hubble
parameter are considered in the Weyl theory of gravity, by taking
them as functions of $r$ and $t$, respectively. Based on this
theory and in the linear approximation, we obtain the values of
$H_0$ and $\Lambda_0$ which are in good agreement with the known
values of the parameters for the current state of the universe.

\end{abstract}

\section{Introduction}
Einstein's general theory of relativity seems to be a perfect
theory (at least in the classical level) which almost all the
classical tests confirm it. This theory is obtained from the
Einstein-Hilbert action
$$I_{EH}=\frac{1}{16\pi G}\int{d^4x}\sqrt{-g}R,$$
where $R$ is the Ricci scalar. However, in some astrophysical
issues, we have to consider other sources of matter and energy
with repulsive gravitational properties (\cite{ja} and Ref.s
therein). For example in the rotation curves of spiral galaxies
where the galactic rotational velocities differ from the
velocities that predicted by the Newtonian gravitational
potentials due to the luminous matter in the galaxies \cite{ja1},
another is related to the origin of the positive accelerating
expansion of our universe. These unseen or dark sources are
presently the most exciting open problems in cosmology and up to
present days there is no satisfactory explanation for the origin
of them. There are various theories to construct acceptable dark
energy models \cite{ja3}, among these models, the modified gravity
approach is attractive, since it tries to explain a natural
gravitational alternative for dark sources. In this theory, the
standard Einstein-Hilbert action is generalized to more
complicated analytic function of the curvature ($f(R)$-gravity).
There exist more fundamental reasons for consideration of such
theories. For example the corrected gravitational potential which
obtained can predict the Milky way rotation curve without the need
of the dark matter scenario. Almost all these theories lead to the
higher order derivative field equations.

In the Weyl theory of gravity the Einstein-Hilbert action is
replaced by the square of the conformal Weyl tensor
\begin{equation}\label{we}
I_W=-\alpha\int{d^4x\sqrt{-g}C_{\mu\nu\rho\lambda}C^{\mu\nu\rho\lambda}},\end{equation}
where $ C_{\mu\nu\rho\lambda}$ is the Weyl tensor
$$C_{\mu\nu\lambda\rho}=R_{\mu\nu\lambda\rho}-\frac{1}{2}\left(g_{\mu\lambda}
R_{\nu\rho}-g_{\mu\rho}R_{\nu\lambda}-g_{\nu\lambda}R_{\mu\rho}+g_{\nu\rho}R_{\mu\lambda}\right)
+\frac{R}{6}\left(g_{\mu\lambda}g_{\nu\rho}-g_{\mu\rho}g_{\nu\lambda}\right).$$
Since there is no accepted system of sign convention in general
relativity, different articles use different sign conventions. In
this work we first review Weyl gravity with the usual sign
convention of Ref. \cite{w,w1}, and then by taking the
cosmological constant and the Hubble parameter as functions of $r$
and $t$ respectively, it is shown that in the Weyl gravity, the
estimated values for these quantities are close to their measured
values. In the appendix, some useful relations have been
presented.

\section{Weyl gravity}

The action (\ref{we}) can be written as follows$$I_W=-\alpha\int
d^4x
\sqrt{-g}(R^{\mu\nu\rho\lambda}R_{\mu\nu\rho\lambda}-2R^{\mu\nu}R_{\mu\nu}+\frac{1}{3}R^2).
$$
Since
$\sqrt{-g}(R^{\mu\nu\rho\lambda}R_{\mu\nu\rho\lambda}-4R^{\mu\nu}R_{\mu\nu}+R^2)$
is a total divergence (Gauss-Bonnet term), it does not contribute
to the equation of motion and one can simplify the action as
follows \cite{Kazanas} \begin{equation} \label{action}
I_W=-2\alpha\int
d^4x\sqrt{-g}\Big(R_{\alpha\beta}R^{\alpha\beta}-\frac{1}{3}R^2\Big)\equiv-2\alpha\int
d^4x \Big({\cal W}_2-\frac{1}{3}{\cal W}_1\Big). \end{equation}
The variation of the action with respect to $g_{\alpha\beta}$
results in
$$\delta({\cal W}_1)=2\sqrt{-g}R\delta
R+R^2\delta\sqrt{-g}$$\begin{equation}\label{var1}
=\sqrt{-g}\nabla_\gamma
A^\gamma+\sqrt{-g}\Big(2\nabla^\alpha\nabla^\beta
R-2g^{\alpha\beta}\Box
R-2RR^{\alpha\beta}+\frac{1}{2}g^{\alpha\beta}R^2\Big)\delta
g_{\alpha\beta},\end{equation} where
$$A^\gamma\equiv2g^{\alpha\beta}g^{\gamma\lambda}R(\nabla_\alpha\delta g_{\beta\lambda}-\nabla_\lambda\delta
g_{\alpha\beta})-2g^{\gamma\beta}g^{\alpha\lambda}\nabla_\alpha
R\delta
g_{\beta\lambda}+2g^{\alpha\beta}g^{\gamma\lambda}\nabla_\lambda
R\delta g_{\alpha\beta},$$ and  $$ \delta({\cal
 W}_2)=\sqrt{-g}\nabla_\gamma B^\gamma+$$\begin{equation}
\label{var2}\sqrt{-g}\Big(\nabla_\rho\nabla^\alpha
R^{\beta\rho}+\nabla_\rho\nabla^\beta R^{\alpha\rho}-\Box
R^{\alpha\beta}-g^{\alpha\beta}\nabla_\rho\nabla_\lambda
R^{\rho\lambda}-2R_\rho^\beta
R^{\rho\alpha}+\frac{1}{2}g^{\alpha\beta}R_{\rho\lambda}R^{\rho\lambda}\Big)\delta
g_{\alpha \beta},\end{equation} where

$$B^\gamma=g^{\gamma \rho}R^{\alpha\beta}\Big(\nabla_\alpha\delta
g_{\beta\rho}+\nabla_\beta\delta g_{\alpha\rho}-\nabla_\rho\delta
g_{\alpha\beta}\Big)-g^{\alpha\rho}\Big(\nabla_\alpha R^{\gamma
\beta}\delta g_{\beta\rho}-\nabla_\beta R^{\gamma \beta}\delta
g_{\alpha\rho}\Big)$$

$$-g^{\beta\rho}\Big(R^{\alpha\gamma}\nabla_\alpha\delta
g_{\beta\rho}+\nabla_\beta R^{\alpha\gamma}\delta
g_{\alpha\rho}\Big)+g^{\rho\gamma}\nabla_\rho
R^{\alpha\beta}\delta g_{\alpha\beta}.$$  When the equation of
motion is considered, the first terms in the right hand side of
Eq.s (\ref{var1}) and (\ref{var2}) have no contribution. Therefore
we obtain
$$W_{\alpha\beta}\equiv W^{(2)}_{\alpha\beta}-\frac{1}{3} W^{(1)}_{\alpha\beta}=\nabla^\rho\nabla_\alpha
R_{\beta\rho}+\nabla^\rho\nabla_\beta R_{\alpha\rho}-\Box
R_{\alpha\beta}-g_{\alpha\beta}\nabla_\rho\nabla_\lambda
R^{\rho\lambda}$$

\begin{equation}\label{m}-2R_{\rho\beta}
R^{\rho}_\alpha+\frac{1}{2}g_{\alpha\beta}R_{\rho\lambda}R^{\rho\lambda}-\frac{1}{3}\Big(2\nabla_\alpha\nabla_\beta
R-2g_{\alpha\beta}\Box
R-2RR_{\alpha\beta}+\frac{1}{2}g_{\alpha\beta}R^2\Big)
.\end{equation} Therefore, $W_{\alpha\beta}=0$ are the vacuum Weyl
field equations\footnote{The relation (\ref{m}) has been written
by taking the following sign convention of \cite{w}
$$R^\alpha_{\beta\gamma\rho}=\partial_\gamma\Gamma^\alpha_{\beta\rho}+\Gamma^\alpha_{\lambda\gamma}\Gamma^\lambda_{\beta\rho}-(\gamma\leftrightarrow
\rho),$$$$R_{\beta\rho}=R^\alpha_{\beta\alpha\rho},\,\,\mbox{and}\,\,\,R=R^\alpha_\alpha,$$
 in Ref. \cite{Kazanas}, the opposite sign for the Ricci tensor and
Ricci scalar has been chosen.}.

In one of the most significant papers (Ref. \cite{mannheim4}), the
authors concern about conformal radial solutions to vacuum Weyl
field equations ($W^{rr}=0$). In the so called article, a general
metric has been derived, covering all spherically symmetric
solutions to Einstein field equations. Moreover, this metric
possesses an additional constant, which is claimed that can
geometrically explain the effects of dark energy. This metric, as
it is described in \cite{mannheim4}, also covers the de Sitter
solution with the cosmological term. In this work we obtain
similar results, by considering de Sitter like metric in the
background field method and linear approximation.

\section{An estimation for $\Lambda$ and $H$}

Let us choose the following metric that mimics the de
Sitter-Schwarzschild metric
\begin{equation}
ds^2 = g_{\alpha\beta}dx^{\alpha}dx^{\beta}=- \left( 1-2\frac
{GM}{r}-\frac{1}{3}f
 \left( r \right) {r}^{2} \right)dt^2 +  \left( 1-2\frac
 {GM}{r}-\frac{1}{3}f
 \left( r \right) {r}^{2} \right)^{-1}dr^2 +
r^2d\Omega^2. \label{arbitrary metric}
\end{equation}
After doing some tedious but straightforward calculation, the
components of the Ricci tensor become as follows:
\begin{equation}
\begin{array}{l}
R_{00}=\frac{1}{2}\Big(1-2\,{\frac {GM}{r}}-\frac{1}{3}\,
{f\,{r}^{2}}\Big)\Big( -\frac{1}{3}\, { f}^{{\it
''}}{r}^{2}-2\, {f}^{'}r-2\, {f} \Big),\\

R_{11}=- \Big(1-2\,{\frac {GM}{r}}-\frac{1}{3}\,
{f\,{r}^{2}}\Big)^{-2}R_{00}, \\

R_{22}= \frac{1}{3}\,{f}^{'} {r}^{3}+\,f\,r^2

  ,\\

R_{33} = R_{22}\left( \sin \left( \theta \right) \right) ^{2}.
\end{array}
\label{RicciSch-ds}
\end{equation}
And the Ricci scalar takes the following form
\begin{equation}
R=\frac{1}{3}\,\Big({f}^{{\it ''}}{r}^{2}+8\,{f}^{'}r+12\,f \Big),
\label{RicciSSch-ds}
\end{equation}
where the prime stands for the derivative with respect to $r$. And
also the component of the Weyl equation is found as

\begin{equation}
\begin{array}{l}
W_{00}=\frac{1}{9}(96\,\frac{(GM)^2}{r^2}{\it
f''}\,r-106\,\frac{GM}{r}{\it f''}\,+\frac{4}{3}\,{{ \it
f'}}^{2}{r}GM+\frac{1}{18}\,{\it f'}\,{r}^{6} f {\it f'''}
+\frac{2}{9}\,{\it f'}\,{r}^{5}  f  {\it f''}
+42\,\frac{(GM)^2}{r}{ \it f'''}\,\\

-\frac{1}{6}\,{r}^{3}{{\it f''}}^{2}GM-43\,GM{\it f'''}-4\,{\it
f''''}\,{r}GM+4\,{(GM)^2}{\it f''''}\,-44\,\frac{GM}{r^2}{\it
f'}\,r +11\,{\it f'''}\,{r}+29\,{\it f''}\,\\

+16\,\frac{{\it f'}}{r}\,-\frac{2}{3}\,{r}^{2}{{\it
f'}}^{2}+\frac{1}{12}\,{r}^{4}{{\it f''}}^{2}+\,{\it
f''''}\,{r}^{2}-\frac{23}{3}\,  f \, {r}^{3}{\it
f'''}+\frac{1}{9}\,{f}^{2}{r}^{6}{\it
f''''}-\frac{1}{6}\,{r}^{4}{\it f'}\,{\it f'''}-\frac{40}{3} \, f
\, {r}{\it
f'}\\

+\frac{2}{9}\,{{\it f'}}^{2}{r}^{4}  f
+\frac{8}{3}\,{f}^{2}{r}^{3}{\it f'}-\frac{2}{3}\,{\it
f''''}\,{r}^{4}  f -\frac{1}{36}{r}^{6}  f \, {{\it
f''}}^{2}-\frac{2}{3}\,{r}^{3}{\it f'} \,{\it
f''}+24\,\frac{(GM)^2}{r^3}{\it f'}+4\,{f}^{2}{r}^{4}{\it
f''} \\

-\frac{65}{3}\,  f \, {r}^{2}{\it f''}+\frac{1}{3}\,{\it
f'}\,{r}^{3}GM{\it f'''}+20\,GM  f \,{\it f'}+\frac{4}{3}\,{\it
f'}\,{r}^{2}GM {\it f''}+40\,GM \,f \, {r}{\it f''}\\

+15\,GM \, f
 \, {r}^{2}{\it f'''}+\frac{4}{3}\,GM{\it f''''}\,{r}^{3} \, f
 +\frac{4}{3}\,{f}^{2}{r}^{5}{\it f'''}),
\end{array}
\label{W00-static}
\end{equation}

$$ W_{11}=\frac{1}{18(
 -3\,r+6\,GM+f\,r^3)}\times
$$

\begin{equation} \Big(-48\,{\it f'}-42\,{\it
f''}\,r-\,{r}^{5}{\it f'}\,{\it f'''}+216\,\frac{GM}{r}{\it
f'}+18\,{\it f'''}\,{r}GM+144\,{\it f''}\,GM-6\,{\it
f'''}\,{r}^{2}+\frac{1}{2}{r}^{5}{{\it f''}}^{2}-4\,{\it f'}\,
{r}^{4}{\it f''}-4\,{{\it f'}}^{2}{r}^{3}\Big), \label{W11-static}
\end{equation}

\begin{equation}
\begin{array}{l}
9 W_{22} = {\frac {1}{12}}\,{r}^{6}{{\it f''}}^{2}-{\frac
{2}{3}}\,{{\it f'}}^{ 2}{r}^{4}+\frac{1}{2}\,{\it
f''''}\,{r}^{4}+5\,{\it f'''}\,{r}^{3}+{ 11}\,{\it
f''}\,{r}^{2}+4\,{\it f'}\,r -4\,  f \, {r }^{3}{\it
f'}+12\,GM{\it
f'}\\

-{\frac {1}{6}}\,{r}^{6}{\it f'}\,{\it f'''}-{\frac {2}{3}}\,{\it
f'}\,{r}^{5}{\it f''}-6\,{\it f''}\,{r}^ {4} f  -{\frac
{1}{6}}\,{\it f''''}\,{r}^{6}  f-2\,{\it f'''}\,{r}^{5} f
-12\,{\it f''}\,r GM -\,{\it
f''''}\,{r}^{3}GM\\

-9{\it f'''}\,{r}^{2}GM\, =\, \frac{9}{\sin^2(\theta)}W_{33}.
\end{array}
\label{W22-static}
\end{equation}
Letting
$W_{\alpha\beta}=0$ results in four differential equations but
here, the common solution for $f(r)$ is considered (the general
form of the solution can be found in \cite{mannheim4})
\begin{equation}
f(r) = c_1+\frac{6GM}{r^3}. \label{vacuum solution}
\end{equation}
Taking into account this value for $f(r)$ in (\ref{arbitrary
metric}), yields
\begin{equation}\label{as} g_{00}=-\Big(1-\frac
{4GM}{r}-\frac{1}{3}c_1 r^2 \Big).\end{equation} This is similar
to the de Sitter-Schwarzschild metric (except the coefficient of
$1/r$) that indicate this metric is a solution as well.

 Now let us use the weak field limit which
implies
\begin{equation}
g_{00}=\eta_{00}+h_{00},\label{metric-rewritten}
\end{equation}
where $\eta_{00}=-1$, and $h_{00}=\frac{ 4GM}{r}+\frac{1}{3}c_1
r^2$ is considered to be very small. The Poisson equation for this
potential implies that
\begin{equation}
\nabla^2 h_{00}=8\pi T_{00},\label{poisson1}
\end{equation}
in which $T_{00}$ is the density of a homogenous spherical
observed mass, $M_{obs}$, with the observed radius, $r_{obs}$.
From (\ref{as}) and (\ref{poisson1}), it follows that
\begin{equation}
\Big(\frac{d^2}{dr^2}+\frac{2}{r}\frac{d}{dr}\Big)\Big(\frac{4GM}{r}+\frac{1}{3}c_1
r^2\Big)=8\pi\rho_{obs} ,\label{poisson2}
\end{equation}
where $\rho_{obs}=\frac{3M_{obs}}{4\pi r^3_{obs}}$. So we find

\begin{equation}
c_1=3\frac{M_{obs}}{r^3_{obs}}. \label{f2}
\end{equation}
This relation may be considered as an estimation for the
cosmological constant (see relation (\ref{as}) with $c_1\equiv
\Lambda$). Taking $r_{ob}\approx4.39\times10^{28}\,cm$
\cite{radius,radius1} and $M_{ob}\approx 8\times10^{55}\,gr$, in
which $r_{obs}$ is the radius of the observable universe and
$M_{obs}$ is the mass amount of a steady-state observable universe
which has been calculated by Sir Fred Hoyle \cite{Hoyle}, we
obtain
$$c_1\equiv \Lambda=3\frac{M_{ob}}{{r_{ob}}^3}\approx 2.84\times10^{-30}\,\frac{gr}{cm^3}.$$
This can be comparable to the value, which has been estimated for
the cosmological constant $\Lambda_0$. Note that the cosmological
constant is measured to be of order $10^{-29}\,\frac{gr}{cm^3}$
\cite{cosmologicalcte.}.\vspace{2mm}

At the end of this section, let us assume a general time dependent
metric as:
\begin{equation}
ds^2=-dt^2+f(t)dx_i^2, \label{General f(t)}
\end{equation}
in which $f(t)$ is an arbitrary function of $t$. It is found that
this general time-dependent metric, is a solution of Eq. (\ref{m})
($W_{\alpha\beta}=0$). Therefore we rewrite the metric
(\ref{General f(t)}) as follows
\begin{equation}
ds^2 = -dt^2 + e^{2h(t)t}dx_i^2. \label{weyl-metric-hubble}
\end{equation} Using this metric, the Ricci
tensor components become as follows:

\begin{equation}
\begin{array}{l}
R_{00} = -3\Big(({\it\ddot h}\,t+2\,{\it \dot h} ) + ( {\it\dot
h}\,t+h ) ^{2}\Big),\\

R_{11}=R_{22}=R_{33} = e^{2ht}\Big(({\it\ddot h}\,t+2\,{\it \dot
h} ) + 3( {\it\dot h}\,t+h ) ^{2}\Big).\label{ricci}
\end{array}
\end{equation}
Also the Ricci scalar is calculated as follows
\begin{equation}
R = 6({\it\ddot h}\,t+2\,{\it\dot h}+2\,{{\it\dot
h}}^{2}{t}^{2}+4\,{\it\dot h}\,t
 h +2\,{h}^{2}), \label{ricci
scalar}
\end{equation} where over dot presents the time derivative. Note that if $h(t)$ is supposed to be a constant then the de Sitter condition will be obtained
\cite{w,w1}.

Using these values in the expressions for $W^{(1)}_{\alpha\beta}$
and $W^{(2)}_{\alpha\beta}$ in Eq. (\ref{m}) gives:
\begin{equation}
\begin{array}{l}
W^{(1)}_{00} = -36\,{\it \dot h}\,t{\it\ddot h}+72\,{{\it\dot
h}}^{2}+18\,{{\it\ddot h}}^{2}{t}^{2 }-36\,{\it\dot
h}\,{t}^{2}{\it h^{(3)}}-36\,{\it h^{(3)}}\,t \, h -
108\,{\it\ddot h}\,  h -216\,{\it\ddot h}\,{t}^{2}{\it\dot h}\,
 h \\

 -216\,{{\it\dot h}}^{3}{t}^{2}-216\,{\it\dot h}\,{h}^{2}-
108\,{\it\ddot h}\,{t}^{3}{{\it\dot h}}^{2}-108\,{\it\ddot
h}\,t{h}^{2}-432\,{{ \it\dot h}}^{2}t \, h
,\\\\

W^{(2)}_{00} = 24\,{{\it\dot h}}^{2}-72{\it\ddot
h}\,{t}^{2}{\it\dot h}\, h-12\,{\it\dot h}\,t{\it\ddot
h}-12\,{\it\dot h}\,{t}^{2}{\it h^{(3)}}-12\,{\it h^{(3)}}\,t \, h
-36\,{\it\ddot h}\,{t}^{3}{{\it\dot h}}^{2}-36\,{\it\ddot
h}\,t{h}^{2}-144\,{{\it\dot h}}^{2}t \, h
 \\

 +6\,{{\it\ddot h}}^{2}{t}^{2}-36\,{\it\ddot h}\, h -
72\,{{\it\dot h}}^{3}{t}^{2}-72\,{\it\dot h}\,{h}^{2}.

\end{array}
\label{W1-W2}
\end{equation}
From the above relations it is clear that $W_{00}
\,(=W^{(2)}_{00}-\frac{1}{3}W^{(1)}_{00})$ becomes zero for all
$h(t)$. Thus let us consider the following $h(t)$
$$h(t)=h_1\frac{\ln t}{t}+h_2,$$
where $h_1$ and $h_2$ are two constants. If we take
$a^2(t)=e^{2h(t)\,t}$, where $a(t)$ stands for the scale factor,
then the Hubble parameter reads as follows $$H=\frac{\dot
a}{a}=\frac{h_1}{t}+h_2.$$ As a simple guess in this work, we take
$h_1=1$ and $h_2=0$. And if $t$ be assumed the age of the universe
- the age of the universe is now calculated to be $13.75\times
10^9 \pm 0.17$ yrs \cite{Hubble} - then one obtains the Hubble
constant for this time as:
\begin{equation}
 H\approx 2.31\times 10^{-18} s^{-1}.
\label{H0}
\end{equation}
The recent measurement for Hubble constant is $H_0 = 73.8 \pm
2.4\,\, (km/s)/Mpc$ or $2.39 \pm 8\% \times 10^{-18}\,\,s^{-1}$
\cite{chandra}.

\section{Conclusion}

Conformal symmetry is indeed one of the most important measures of
assessment of massless field in quantum field theory. If the
graviton does exist and propagates on the light cone due to its
long range effect, it should have zero mass. The light cone
propagation immediately imposes the conformal invariance on the
graviton field equations. Historically, the first conformally
invariant theory of gravity introduced by Weyl in 1918; then
Einstein pointed out that the non-integrability of the lengths of
vectors under Weyl-like parallel propagation contradicts to
physical experience (very good bibliography and background are
given in \cite{HWE}). In recent years there is a renewal interest
of such scale-invariant theories. For example Friedmann equation
of cosmological evolution for this theory can be found in
\cite{fri,fri1}. In Weyl gravity, the usual Einstein-Hilbert
action, is replaced by the square of the conformal Weyl tensor.
This leads to a gravitational theory of fourth order. Previously,
exact vacuum solution to conformal Weyl gravity had been studied
in \cite{mannheim4}, however in this paper, it has been shown that
some significant radial solutions, (such as de Sitter and
Reissner-Nordstr\"{o}m metrics), in the linear approximation and
the background field method, can be easily obtained. We then
studied the time and $r$ dependence of the Hubble and cosmological
parameters in the Weyl theory of gravity. As a result, one may
obtain theoretically some observable quantities, with high
accuracy, in higher order theories of gravity (including the Weyl
theory), as it has been confirmed by other authors
\cite{tekin,tekin1}. \vspace{2mm}

\noindent {\bf{Acknowledgments}}:  We would like to thank H.
Ardahali and S. Fatemi for their interest in this work. We also
thank the referee for his/her useful comments.

\appendix

\section{appendix } Recent astrophysical data indicate that our
universe has a positive and non-vanishing acceleration, therefore
it might well be explained by de Sitter model. De Sitter
space-time can be considered as a vacuum solution of the Weyl
gravity. Considering the condition
\begin{equation} \label{co}R_{\mu\nu}=\frac{1}{4}g_{\mu\nu}R,\end{equation} together with the
Bianchi identity, one obtains $\nabla_\mu R=0$, it means that $R$
is a constant. Imposing this condition into the Einstein's
equation, leads to
 \begin{equation}
R=4\Lambda. \label{trace-einstein}
\end{equation}
This is the well-known de Sitter condition, which is followed by a
constant $R$. De Sitter space-time is maximally symmetric and we
have \cite{w,w1},

$$
R_{abcd}=H^2(g_{ac}g_{bd}-g_{ad}g_{bc}),\,\,R_{ab}=3H^2g_{ab},\,\,R=12H^2=4\Lambda.$$
The following relations become important in writing the Weyl
gravity in de Sitter background and in the linear form $$\delta
\Gamma^a_{bd}=\frac{1}{2}g^{ae}\Big(\nabla_b\delta
g_{ed}+\nabla_d\delta g_{be}-\nabla_e\delta g_{bd}\Big),$$$$\delta
R^a_{bcd}=\frac{1}{2}g^{ae}\Big(\nabla_c\nabla_b\delta
g_{de}+\nabla_c\nabla_d\delta g_{be}-\nabla_c\nabla_e\delta
g_{bd}-\nabla_d\nabla_b\delta g_{ce}-\nabla_d\nabla_c\delta
g_{be}+\nabla_d\nabla_e\delta g_{bc}\Big),$$ $$\delta
R_{ab}=\frac{1}{2}g^{cd}\Big(\nabla_c\nabla_a\delta
g_{bd}+\nabla_c\nabla_b\delta g_{ad}-\nabla_c\nabla_d\delta
g_{ab}-\nabla_b\nabla_a\delta g_{cd}\Big),$$$$\delta
R=\Big(\nabla^a\nabla^b\delta g_{ab}-g^{ab}\Box \delta
g_{ab}\Big)-3H^2 g^{ab}\delta g_{ab}.$$

\end{document}